\documentclass[12pt,preprint]{aastex}

\newcommand{\etal}{et~al.\ }

\newcommand{\cmsq}{\hbox{cm$^{-2}$}}

\newcommand{\nh}{\hbox{${N}_{\rm H}$}}

\newcommand{\chandra}{{\emph{Chandra}}}

\newcommand{\asca}{{\emph{ASCA}}}

\newcommand{\mg} {MG~0414+0534}
\newcommand{\he} {HE~1104$-$1805}
\newcommand{\pks} {PKS~1830$-$211}
\newcommand{\sbs} {SBS~0909+523}
\newcommand{\fbqs} {FBQS~0951+2635}
\newcommand{\bone} {B~1152+199}

\begin{document}

\def\sarc{$^{\prime\prime}\!\!.$}
\def\arcsec{$^{\prime\prime}$}
\def\arcmin{$^{\prime}$}
\def\degr{$^{\circ}$}
\def\seco{$^{\rm s}\!\!.$}
\def\ls{\lower 2pt \hbox{$\;\scriptscriptstyle \buildrel<\over\sim\;$}} 
\def\gs{\lower 2pt \hbox{$\;\scriptscriptstyle \buildrel>\over\sim\;$}} 
 
\title{Differential X-ray Absorption and Dust-To-Gas Ratios of the Lens Galaxies SBS~0909+523, FBQS~0951+2635, and B~1152+199}

\author{Xinyu Dai\altaffilmark{1} and Christopher S. Kochanek\altaffilmark{1}}

\altaffiltext{1}{Department of Astronomy,
The Ohio State University, Columbus, OH 43210,
xinyu@astronomy.ohio-state.edu, ckochanek@astronomy.ohio-state.edu}

\begin{abstract}
We analyzed \chandra\ observations of three gravitational lenses, SBS~0909+523, FBQS~0951+2635, and B~1152+199, to measure the differential X-ray absorption and the dust-to-gas ratio of the lens galaxies.
We successfully detected the differential X-ray absorption in SBS~0909+523 and B~1152+199, and failed to detect it in FBQS~0951+2635 due to the dramatic drop in its flux from the \emph{ROSAT} epoch.
These measurements significantly increase the sample of dust-to-gas ratio 
measurements in cosmologically-distant, normal galaxies.
Using the larger sample, we obtain an average dust-to-gas ratio of $E(B-V)/$\nh $=(1.5\pm0.5)\times$ 10$^{-22}$ mag cm$^2$ atoms$^{-1}$ 
with an estimated intrinsic dispersion in the ratio of $\simeq $40\%.  
This average dust-to-gas ratio is consistent with our previous measurement, and 
the average Galactic value of $1.7 \times 10^{-22}$ mag cm$^2$ atoms$^{-1}$ and the estimated intrinsic dispersion is also consistent with the 30\% observed in the Galaxy.  
A larger sample size is still needed to improve the measurements and to
begin studying the evolution in the ratio with cosmic time.
We also detected X-ray microlensing in SBS~0909+523 and significant X-ray variability in FBQS~0951+2635.
\end{abstract}


\section{Introduction}
The inter-stellar medium (ISM) and inter-galactic medium (IGM) are ubiquitous, and understanding the ISM is crucial to many areas of astronomy such as cosmology (e.g., using SNe Ia or GRBs as cosmological probes), the evolution of galaxies, and star formation (e.g., Fall et al. 1996; Madau et al. 1998; Schneider et al. 2004; Corasaniti 2006; Ghirlanda et al. 2006; Lacey et al. 2008).  
The dust-to-gas ratio is a basic property of the ISM, and it is difficult to measure in
  distant galaxies.  In the Galaxy, it is measured by comparing
  the optical extinction and Ly$_\alpha$ absorption towards stars, and a typical
  value is $E(B-V)/\nh = 1.7\times10^{-22}$ mag cm$^{2}$ atoms$^{-1}$ with a scatter about the mean of
  order 30\% \citep{bsd78}. 
Subsequent studies have extended these measurements to additional lines of sight in our Galaxy and to the LMC and SMC, but the basic results are little altered aside from some evidence that the dust-to-gas ratios of the LMC and SMC are modestly lower ($\sim 30$\%) than in the Galaxy (see the review by Draine 2003 and references therein).
This absorption method is hard to apply to extragalactic sources
  where we lack precise knowledge of the intrinsic source
  spectra and tend to be probing low density parts of the
  IGM using absorption lines in the spectra of quasars or $\gamma$-ray
  bursts (e.g., Kacprzak et al. 2008) rather than the inner regions of normal
  galaxies.  An alternate approach is to use emission by
  the dust in the far-infrared to estimate a dust mass and
  then compare it to the estimated gas mass (e.g., Contini \& Contini 2007).

Gravitational lensing provides a means of studying the properties of the ISM
through differential absorption in the lens galaxies between the locations of the quasar images.
By studying the \emph{differences} in the absorption (or extinction) between multiple images, any contamination from the Galaxy in
the foreground (Milky Way) or the quasar host galaxy in the background is essentially eliminated and we can be confident that we are probing the ISM of the lens galaxy.
Moreover, since we are comparing two images of the same quasar, the method has many of the quantitative advantages of the local approach using stars of known spectral type.
In optical bands, this technique has been widely used to study dust extinction and the dust extinction law in lens galaxies \citep[e.g.,][]{na91,fal99,thb00,mo02,wu03,mu04,me05,ela06}.
In X-rays, \citet{da03} and \citet{dk05} measured the differential column densities in two lenses Q~2237+0305 and B~1600+434.  Subsequently,
\citet{dai06} expanded the results to four lenses and proposed that the evolution of the 
dust-to-gas ratio can be determined given a larger sample.  
In this paper, we present new measurements based on 
\chandra\ observations of the quasar lenses \sbs\ (Kochanek et al.\ 1997), \fbqs\ (Schechter et al.\ 1998), and \bone\ (Myers et al.\ 1999).

\section{Observations and Data Reduction}
\begin{figure}
\epsscale{1}
\plotone{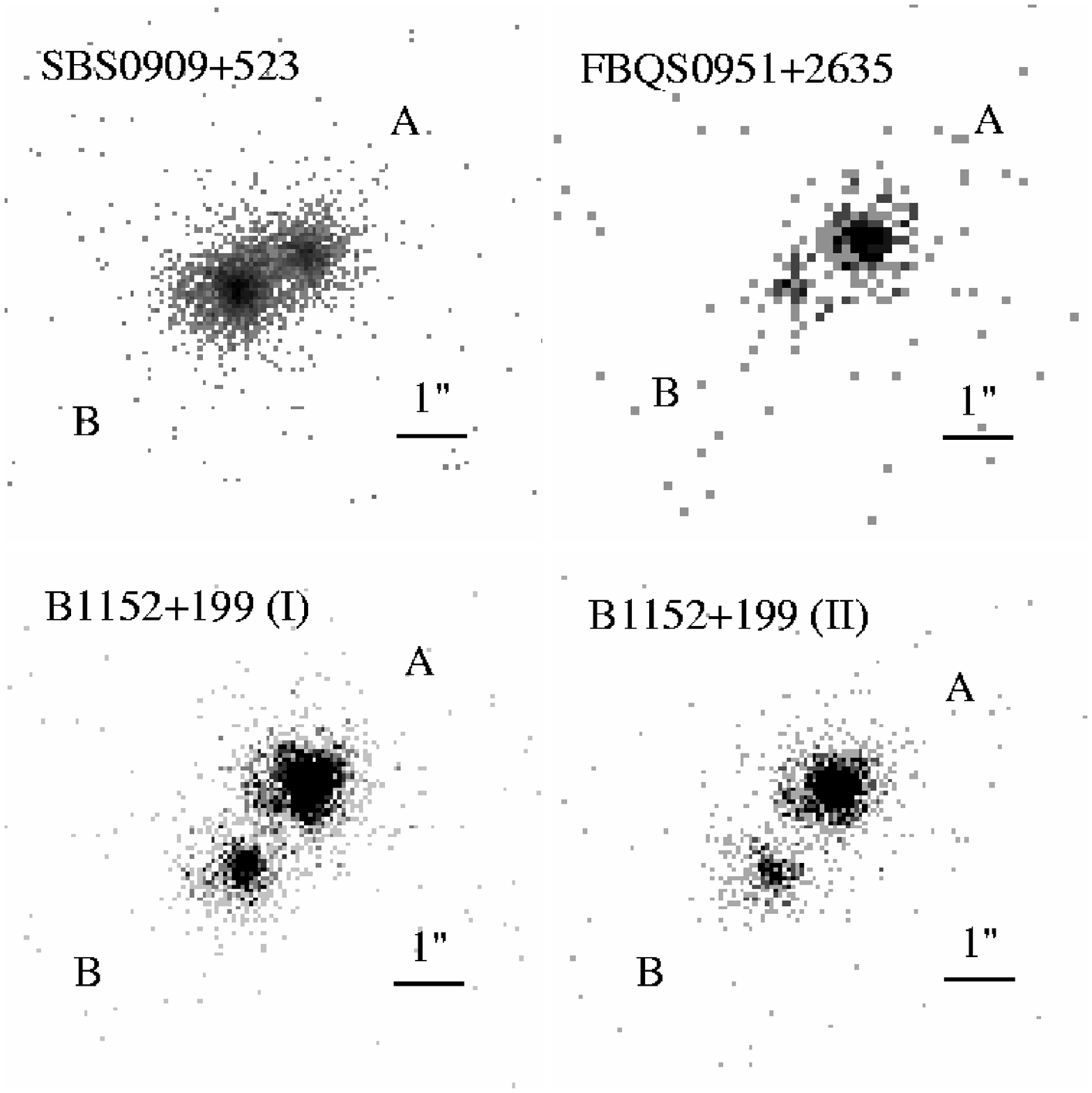}
\caption{\chandra\ images of \sbs, \fbqs, and \bone. \label{fig:img}}
\end{figure}

We observed \sbs\ and \fbqs\ with the Advanced CCD Imaging Spectrometer \citep[ACIS,][]{g03} on board \chandra\ \citep{we02} for 19.6 and 34.2 ks on December 17, 2006 and March 24, 2007, respectively.
\bone\ was observed in a separate program (PI: K. Pedersen) with \chandra-ACIS for 16.4 and 8.3 ks on February 22, 2005 and February 26, 2005.
The back-side illuminated S3 chip was used in all the observations.
A journal of the \chandra\ observations is presented in Table~\ref{tab:obs}.
The \chandra\ data were reduced using the \verb+CIAO 4.0+ software tools provided by the \chandra\ X-ray Center (CXC), following the standard threads on the CXC website.\footnote{The CXC website is at http://cxc.harvard.edu/.}  
We used the most recently reprocessed data products (Reprocessing III) and calibration files (CALDB 3.4.2).
Only events with standard \asca\ grades of 0, 2, 3, 4, and 6 were used in the analysis.  We improved the image quality of the data by removing the pixel
randomization applied to the event positions by the standard pipeline.
In addition, we applied a subpixel resolution technique \citep{t01,m01} to the events on the S3 chip of ACIS where the lensed images are located.
Figure~\ref{fig:img} shows the resulting images.

\begin{deluxetable}{ccccc}
\tabletypesize{\scriptsize}
\tablecolumns{5}
\tablewidth{0pt}
\tablecaption{\chandra\ Observations of Gravitational Lenses \sbs, \fbqs, and \bone. \label{tab:obs}}
\tablehead{
\colhead{Lenses} &
\colhead{Date} &
\colhead{Exp (sec)} &
\colhead{Count Rate (ct/s, 0.3--8 keV)} &
\colhead{Raw Flux Ratio (A/B)\tablenotemark{a}} 
}

\startdata
\sbs & 2006-12-17 & 19654 & $0.162\pm0.003$ & $0.34\pm0.02$ \\
\fbqs & 2007-03-24 & 34243 & $0.0074\pm0.0005$ & $5.2\pm1.4$ \\
\bone & 2005-02-22 & 16449 & $0.189\pm0.003$ & $3.06\pm0.14$ \\
\bone & 2005-02-26 & 8282 & $0.198\pm0.005$ & $3.83\pm0.26$ \\
\enddata

\tablenotetext{a}{The raw flux ratio has not been corrected for the differential absorption
between the two images.}
\end{deluxetable}

\section{Spectral Analysis} 
We fit the spectra of the lensed quasars using \verb+XSPEC V11.3.1+ \citep{a96} over the 0.3--8 keV observed energy range.
The X-ray spectrum of the $i$th image was modeled as
\begin{eqnarray}
N_i(E, t) & = & N_{0, i}(t-\Delta t_i)\left(\frac{E}{E_0}\right)^{-\Gamma(t-\Delta t_i)} \nonumber\\
& & \exp\left\{-\sigma(E)\nh_{,Gal}-\sigma\left[E(1+z_l)\right]\nh_{,i}-\sigma\left[E(1+z_s)\right]\nh_{,Src}(t-\Delta t_i)\right\},
\end{eqnarray}
where $N_i(E, t)$ is the number of photons per unit energy interval, $\nh_{, Gal}$, $\nh_{,i}$, and $\nh_{,Src}$ are the equivalent hydrogen column densities in our Galaxy, the lens galaxy at the position of image $i$, and in the source galaxy, respectively, $\sigma(E)$ is the photo-electric absorption cross-section, and $\Delta t_i$ is the time-delay. 
Although, absorption by the Galaxy, lens and source is degenerate when fitting the
spectrum of each image, the differential absorption $\Delta\nh$ between the lensed images 
is not and can be determined from a simultaneous fit to the spectra of the images.
The time-delay at the source complicates the measurement, since if the absorption at source
is variable (e.g., \pks, Dai et al.\ 2008), it will not be exactly canceled when comparing the two spectra.
However, since the photo-electric absorption cross-section decreases
with energy $\sigma(E) \propto E^{-3}$, the absorption at the source will affect the 
spectrum less.
Thus, we neglected the time-delay between the images and the absorption at the source redshift in this analysis.
We note that a significant change in the spectral index on the time scale
of the time delays will affect the
results, especially for quasars with poor S/N spectra where the spectral index is degenerate with the absorption.  The problem is less severe for
quasars with moderate to high S/N spectra where the spectral index is less
degenerate with \nh\ absorption during spectral modeling.

A second systematic error we must consider is microlensing of the quasar
by the stars in the lens galaxy.  Our analysis is only affected by chromatic
microlensing, where there are significant changes in the
optical and X-ray spectra that alter the estimated extinction or
absorption.   Indeed, it needs only be achromatic ``locally'' to 
separate optical and X-ray wavelength/energy regimes, as global
shifts of the optical spectrum relative to the X-ray spectrum
by microlensing also have no effect on the results.  For the
systems we consider, we have only limited control on the level of
microlensing.   While X-ray microlensing has been detected 
(Chartas et al.\ 2002, 2007; Dai et al.\ 2003; Blackburne et al. 2006; 
Pooley et al. 2007),
X-ray spectral changes due to microlensing have yet to be detected in large
part due to the limited sensitivity of the data.  Chromatic
microlensing in the optical has been observed (Poindexter et al.\ 2007; 
Anguita et al.\ 2008;
Eigenbrod et al.\ 2008) and can create systematic problems for estimates of
the extinction as we had already discovered for \he\ (Dai et al.\ 2006).
We can, however, compare the optical and X-ray flux ratios, since
the evidence to date is that the X-ray sources are much more compact
than the optical (Kochanek et al.\ 2006; Morgan et al.\ 2008; Chartas et al.\ 2008) and so more strongly affected by microlensing.
If the extinction and absorption-corrected flux ratios of the images
agree, this is a good indication that our estimates are little affected by
microlensing.

\begin{figure}
\epsscale{1}
\plotone{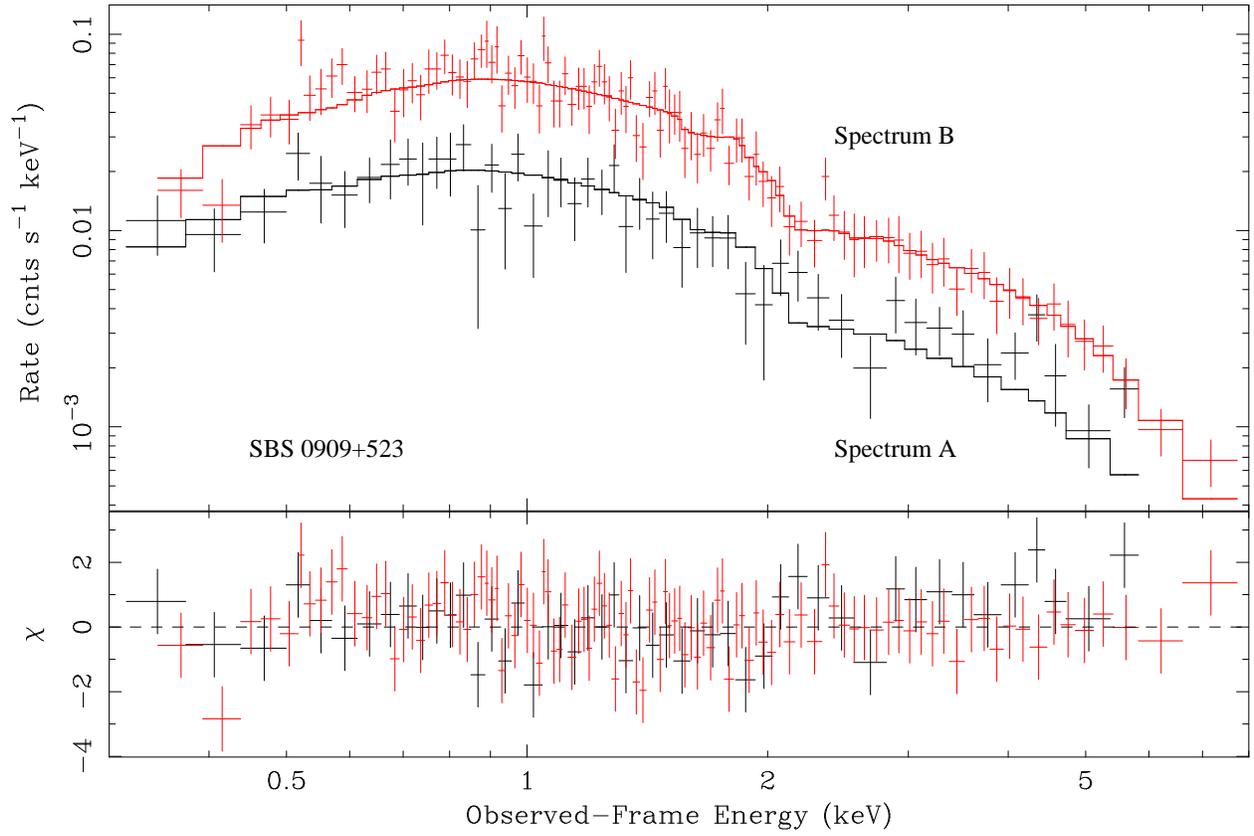}
\caption{Spectra of images A (bottom) and B (top) of \sbs. The lower panel shows the 
contributions of the residuals to the $\chi^2$ fit statistics. \label{fig:sbs}}
\end{figure}

\begin{figure}
\epsscale{1}
\plotone{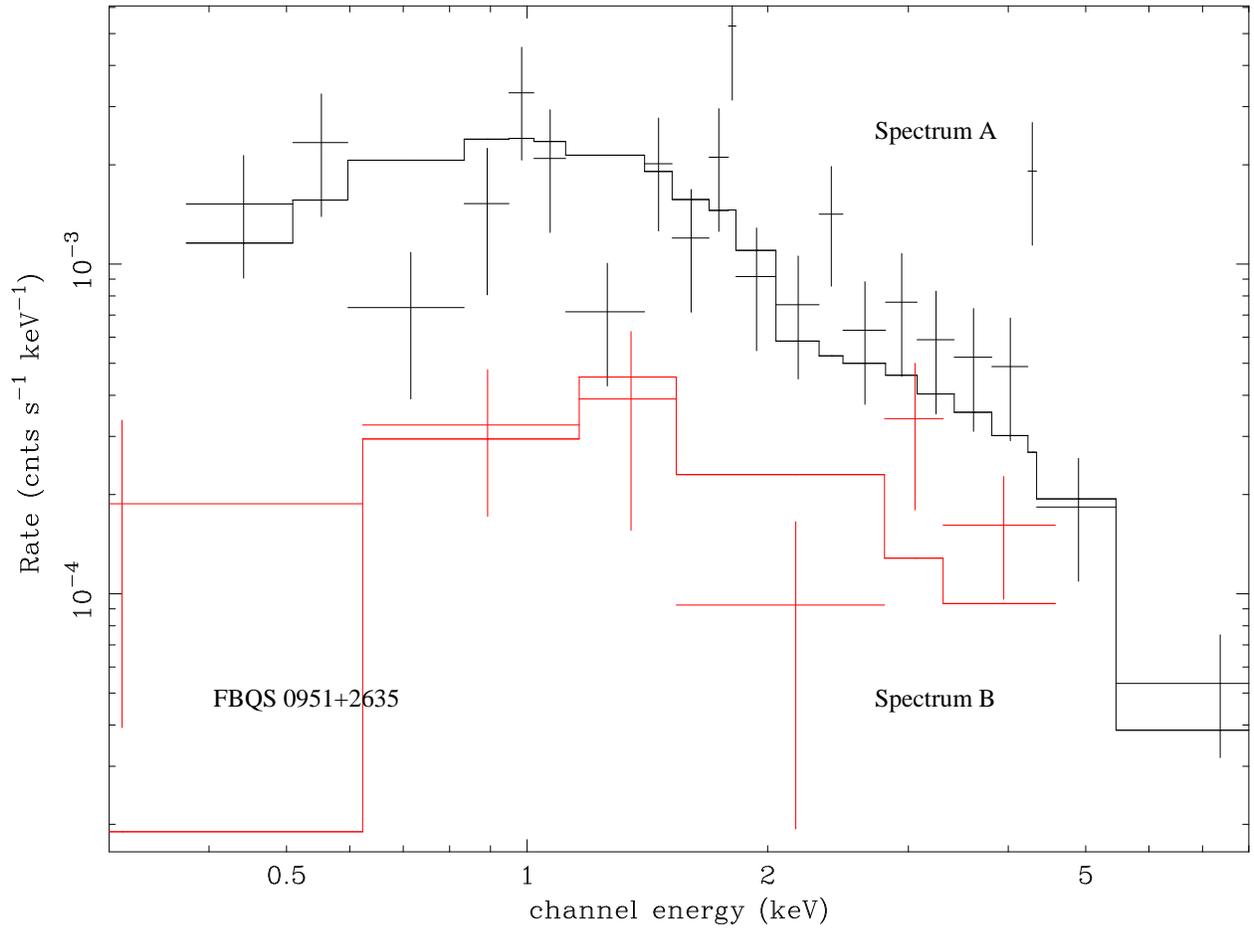}
\caption{Spectra of images A (top) and B (bottom) of \fbqs. \label{fig:fbqs}}
\end{figure}

We used the standard \verb+wabs+ and \verb+zwabs+ models in \verb+XSPEC+ to model the Galactic absorption and absorption at the lens redshift.  The \verb+wabs+ and \verb+zwabs+ models use cross sections from \citet{mm83} and assume a solar elemental abundance from \citet{ae82}.  The Galactic \nh\ is fixed using the values from \citet{d90}.
We fixed the intrinsic power-law index $\Gamma$ to be the same for both lensed images, and allowed the absorption at the lens to differ.
Since the average lens absorption and the power-law index are correlated in the spectral modeling, the uncertainties in absorption by the lens for the individual images are correlated.  
To accurately measure the uncertainties in the differential absorption, we used \verb+XSPEC+ to explore the full parameter space of the lens column densities and the power-law index.  The fitting results are listed in Table~\ref{tab:spec}, and the spectra and best fit models are shown in Figures~\ref{fig:sbs}, \ref{fig:fbqs}, and \ref{fig:bone}.  
We also obtained the unabsorbed flux ratios between the quasar images as part of the model, and 
verified that adding 
absorption in the source does not alter the estimates of the differential column densities.
We describe the results for each of the three lenses in the following sub-sections.

\subsection{\sbs}
\sbs\ (Kochanek et al.\ 1997; Oscoz et al.\ 1997) is a two-image lens system with a source redshift of $z_s = 1.377$ and a lens redshift of $z_l = 0.83$.    
\sbs\ is well studied in the optical bands, and the extinction curve has been measured
using the differential extinction method (Mediavilla et al.\ 2005) to be a Galactic 
extinction curve with a strong 2175\AA\ extinction feature (Motta et al.\ 2002).

It is clear from the \chandra\ spectra that image B is more heavily absorbed than image A.  We estimate that the difference in the 
column densities is $\Delta\nh_{B,A} = (0.055^{+0.095}_{-0.022}) \times 10^{22}$~\cmsq.   
There are several differential extinction measurements for this system (Falco et al.\ 1999; Motta et al. 2002; Mediavilla et al.\ 2005), and we used the most
recent measurement, $\Delta E(B-V) = 0.32\pm0.01$, from Mediavilla et al.\ (2005). 
Combined with our differential absorption measurement, we obtained a dust-to-gas ratio of 
$E(B-V)/\nh = (5.8^{+3.9}_{-3.7})\times10^{-22}$ mag cm$^2$ atoms$^{-1}$. 

The optical flux ratio of \sbs\ has changed little from its discovery
in 1997 through 2006 (Kochanek et al.\ 1997; Motta et al.\ 2002; 
Goicoechea et al.\ 2008)
suggesting that there is little microlensing variability even if there
is evidence for microlensing from the difference between the continuum
and emission line flux ratios (Motta et al.\ 2002; Mediavilla et al.\ 2005). 
The extinction and
absorption corrected optical ($A/B=0.35\pm0.02$, Mediavilla et al.\ 2005) 
and X-ray flux ratios ($A/B=0.32\pm0.03$) 
are consistent but
differ from the $H$-band ($A/B=1.12$, Lehar et al.\ 2000) and extinction-corrected 
emission line flux ratios (Mediavilla et al.\ 2005), indicating that
there is microlensing in this system but it is affecting the optical
and X-ray similarly.  The lack of variability and the similar optical/X-ray
flux ratios suggests the lens is in a region of
relatively uniform microlensing magnification.  We also note that
the X-ray variability of the source is modest. 
\sbs\ was observed three times by \emph{ROSAT}, and the source showed moderate 
X-ray variability of about 40\% (Chartas 2000).
The flux observed by \chandra\ is in between the three \emph{ROSAT} flux measurements,
suggesting no significant total X-ray variability.

\begin{deluxetable}{cccccccc}
\tabletypesize{\scriptsize}
\rotate
\tablecolumns{8}
\tablewidth{0pt}
\tablecaption{Spectral Fitting Results \label{tab:spec}}
\tablehead{
\colhead{} &
\colhead{} &
\colhead{Galactic \nh} &
\colhead{\nh\ (B $-$ A)} &
\colhead{} &
\colhead{} &
\colhead{} &
\colhead{}
\\
\colhead{Quasar} &
\colhead{Model\tablenotemark{a}} &
\colhead{($10^{22}$~\cmsq)} &
\colhead{($10^{22}$~\cmsq)} &
\colhead{$\Gamma$} &
\colhead{Flux Ratio (A/B)} &
\colhead{$\chi^{2}$(dof)} &
\colhead{C-Stat(dof)} 
}

\startdata
\sbs\ & wabs(zwabs(pow)) & 0.0169 (fixed) & $0.055_{-0.022}^{+0.095}$ & $1.63\pm0.05$ & $0.32\pm0.03$ & 124.5(148) & \nodata \\
\fbqs\ & wabs(zwabs(pow)) & 0.0229 (fixed) & $0.49_{-0.41}^{+0.49}$ & $1.32$ (fixed)\tablenotemark{b} & $3.5\pm1.6$ & \nodata & 56.9(29) \\ 
\bone\ (Epoch I) & wabs(zwabs(pow)) & 0.0273(fixed) & $0.48\pm0.06$ & $2.08\pm0.05$ & $1.87\pm0.14$ & 148.23(136) & \nodata \\
\bone\ (Epoch II) & wabs(zwabs(pow)) & 0.0273(fixed) & $0.48\pm0.08$ & $2.10\pm0.05$ & $2.28\pm0.25$ & 54.8(75) & \nodata \\
\bone\ (Combined) & wabs(zwabs(pow)) & 0.0273(fixed) & $0.48\pm0.04$ & $2.07\pm0.03$ & \nodata & 213.1(217) & \nodata \\
\enddata

\tablecomments{Simultaneous fits to the individual spectra of the lensed quasars.  The spectra are constrained to have the same intrinsic power-law photon index but can have different absorption column densities at the redshift of the lens.}
\tablenotetext{a} {Wabs, zwabs, and pow are the XSPEC models used for the Galactic absorption, absorption at the lens, and the intrinsic power-law spectrum, respectively.}
\tablenotetext{b} {The photon index is fix to the best fit value performed in the 2--8 keV.}
\end{deluxetable}

\subsection{\fbqs}
\fbqs\ is a two-image lens discovered by Schechter et al. (1998) with a source redshift of $z_s = 1.24$ and a lens redshift of $z_l = 0.26$ (Eigenbrod et al.\ 2007). 
\fbqs\ was detected in the \emph{ROSAT} All Sky Survey (RASS) with a count rate of 0.02 ct~s$^{-1}$ in the 0.1--2.4 keV band (Voges et al.\ 2000).
Our \chandra\ image shows that \fbqs\ is the brightest source within a 2\arcmin\ radius circle, which indicates that the source detected in RASS is \fbqs.
However, the \chandra\ count rate of \fbqs, 0.007 ct~s$^{-1}$ in the 0.3--8 keV band, is significantly lower than the predicted count rate (0.06 ct~s$^{-1}$) from the RASS, suggesting significant source variability.

The unexpected drop in flux resulted in
low S/N spectra that cause significant challenges in measuring the differential absorption. 
We used the C-statistic (Cash 1979), suitable for small number statistics, to fit the spectra.
Using our standard technique, we found a power-law index of $\Gamma=0.81\pm0.15$ for 
the source.  
We failed to detect the differential absorption in this lens, as the differential absorption
between the two images is consistent with zero.
Note, however, that the upper limit for the lens absorption in image B, $\nh_B < 0.22 \times 10^{22}$~\cmsq\ is larger than that for image A, $\nh_A < 0.05 \times 10^{22}$~\cmsq, 
and that the best fit power-law index ($\Gamma=0.81\pm0.15$) is unusually flat (e.g., Reeves \& Turner 2000; Saez et al.\ 2008).
 A more plausible interpretation is that the low energy part of the spectrum is absorbed.  
Since the absorption is correlated with the
power-law index in the spectral fitting, which is a severe problem for low S/N spectra,
as an experiment we first fit the spectra in the hard band (2--8 keV) to find
a steeper, and more sensible, intrinsic power-law ($\Gamma=1.32$).
Then we held this fixed and fit for the absorption to find a differential
absorption of
$\Delta\nh_{B,A} = (0.49^{+0.49}_{-0.41}) \times 10^{22}$~\cmsq.
Due to the low significance of this detection (or non-detection at all) because of the low S/N spectra, we will be unable to include this lens
in an estimate of the mean dust-to-gas ratio.

\begin{figure}
\epsscale{1}
\plotone{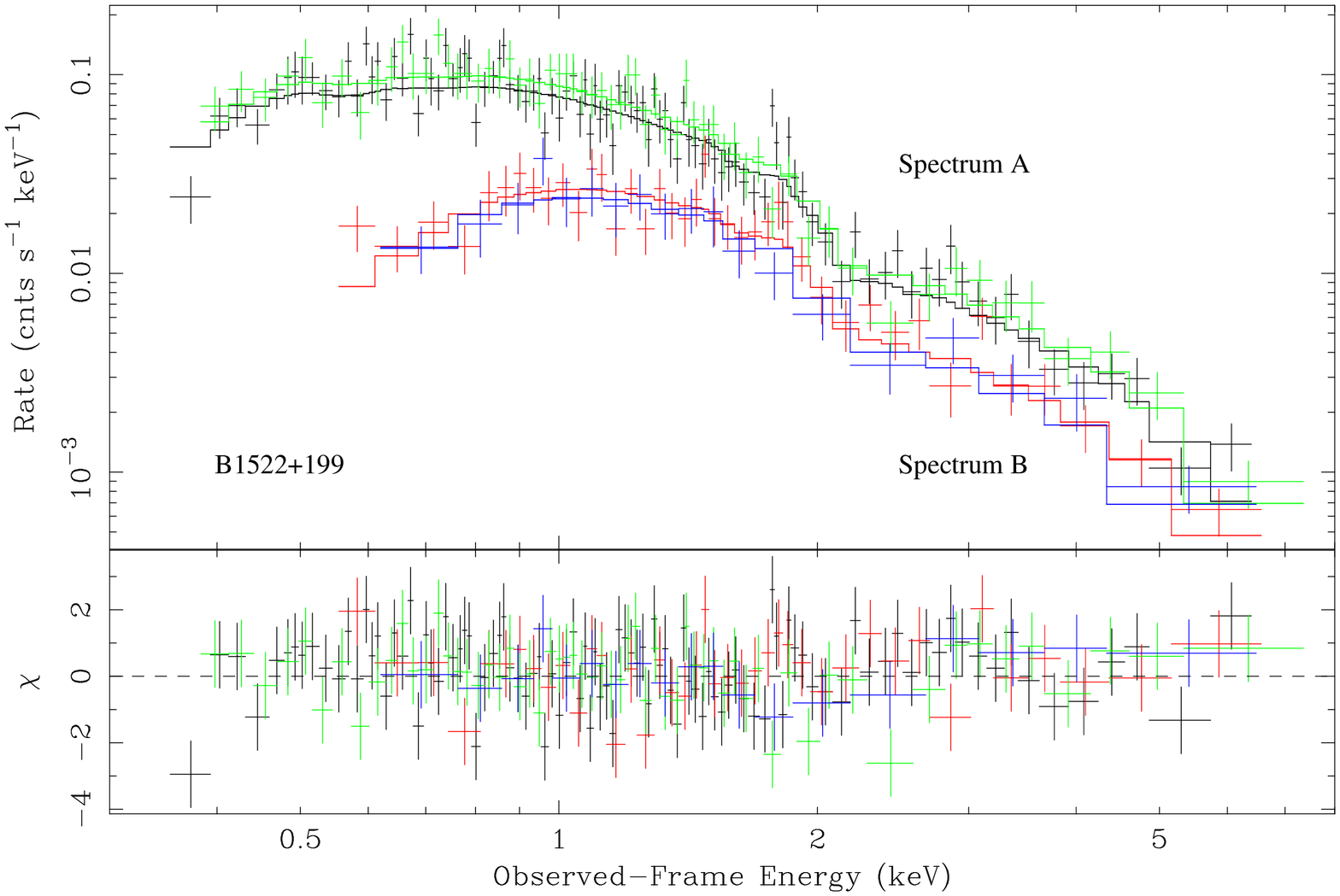}
\caption{Spectra of images A (top) and B (bottom) of \bone\ for both \chandra\ observations.  The black and red lines are the fits for the spectra A and B in the first \chandra\ 
epoch, and the green and blue lines are for the second epoch.  The lower panel shows the
contributions of the residuals to the $\chi^2$ fit statistics.\label{fig:bone}}
\end{figure}

\subsection{\bone}
\bone\ (Myers et al.\ 1999) consists of two $z_s = 1.019$ quasar images lensed by a $z_l = 0.439$ galaxy.  
The dust extinction curve of \bone\ was measured to be $1.3 \le R_V \le 2.0$ (Toft et al.\ 2000) and $R_V=2.1\pm0.1$ (El{\'{\i}}asd{\'o}ttir et al.\ 2006).
The El{\'{\i}}asd{\'o}ttir et al.\ (2006) results were based on improved data
so we adopt their results.
The differential absorption can be clearly seen by comparing the \chandra\ spectra of the two images (Figure~\ref{fig:bone}).
Since \bone\ was observed by \chandra\ twice, we used our standard method and simultaneously
 fit the spectra for both epochs.
We detected a large differential absorption $\Delta\nh_{B,A} = (0.48\pm0.04)\times 10^{22}$~\cmsq.
We also fit the two \chandra\ epochs separately, and obtained consistent
differential absorptions for both observations (Table~\ref{tab:spec}).
We used the model-averaged extinction of $E(B-V) = 1.20\pm0.05$ mag from El{\'{\i}}asd{\'o}ttir et al.\ (2006), where the small scatter in the value
between their models would have little effect on our results.
Combining the differential extinction and absorption measurements, we found a dust-to-gas ratio for the lens galaxy of $E(B-V)/\nh = (2.5\pm0.2)\times10^{-22}$ mag cm$^2$ atoms$^{-1}$.  This is the most accurate dust-to-gas ratio measurement we have obtained so far.  
There is weak evidence ($1.4\sigma$) for a change in the X-ray flux ratios, where the flux ratio (A/B) is larger in the second epoch (see Table~\ref{tab:spec}).
The X-ray flux ratios after correcting for absorption ($A/B = 1.87\pm0.14$ and $2.28\pm0.25$) are also consistent with the range of optical flux ratios after correcting for extinction (e.g., $\Delta M = 0.85\pm0.07$, $0.6\pm0.1$, El{\'{\i}}asd{\'o}ttir et al.\ 2006).

\begin{deluxetable}{ccccccc}
\tabletypesize{\scriptsize}
\tablecolumns{7}
\tablewidth{0pt}
\tablecaption{The Dust-To-Gas Ratio of High Redshift ($z>0$) Galaxies \label{tab:dtg}}
\tablehead{
\colhead{} &
\colhead{} &
\colhead{} &
\colhead{} &
\colhead{$\Delta \nh$} &
\colhead{$\Delta E(B-V)$} &
\colhead{$E(B-V)/\nh$} 
\\
\colhead{Lens} &
\colhead{$z_l$} &
\colhead{type} &
\colhead{Between Images} &
\colhead{($10^{22}$~\cmsq)} &
\colhead{(mag)} &
\colhead{($10^{-22}$ mag cm$^2$ atoms$^{-1}$)} 
}
\startdata
\cutinhead{New Measurements}
\sbs      & 0.83  & elliptical & B, A  & $0.055^{+0.095}_{-0.022}$ & $0.32\pm0.01$ & $5.8^{+3.9}_{-3.7}$ \\
\fbqs     & 0.26  & elliptical & B, A  & $0.49^{+0.49}_{-0.41}$ & $-0.12\pm0.02$ & \nodata \tablenotemark{a} \\
\bone     & 0.439 & elliptical & B, A  & $0.48\pm0.04$ & $1.20\pm0.05$ & $2.5\pm0.2$ \\
\cutinhead{Dai et al.\ (2006) Sample}
\mg       & 0.9584 & elliptical & A, B & $0.33\pm0.10$          & $0.18\pm0.11$ & $0.55\pm0.33$ \\   
\he        & 0.729 & elliptical & A, B & $0.055\pm0.030$        & $-0.07\pm0.01$          & \nodata \tablenotemark{a} \\
B~1600+434  & 0.41   & spiral & B, A & $0.26^{+0.17}_{-0.12}$ & \phs$0.10\pm0.03$           & \phd$0.38\pm0.25$  \\
\pks       & 0.886  & spiral & B, A & $1.8^{+0.5}_{-0.6}$    & \phs$3.00\pm0.13$           & $1.7\pm0.6$ \\
Q~2237+0305 & 0.0395 & spiral\tablenotemark{b} & A, C & $0.04\pm0.03$          & \phs$0.11\pm0.03$           & $2.8\pm2.2$ \\
\enddata

\tablenotetext{a} {The dust-to-gas ratio was not estimated for \fbqs\ and \he\ because the differential extinction and \nh\ column measurements give opposite signs.}
\tablenotetext{b} {Q~2237+0305 is lensed by the central bulge of a spiral galaxy.}
\end{deluxetable}

\begin{figure}
\epsscale{1}
\plotone{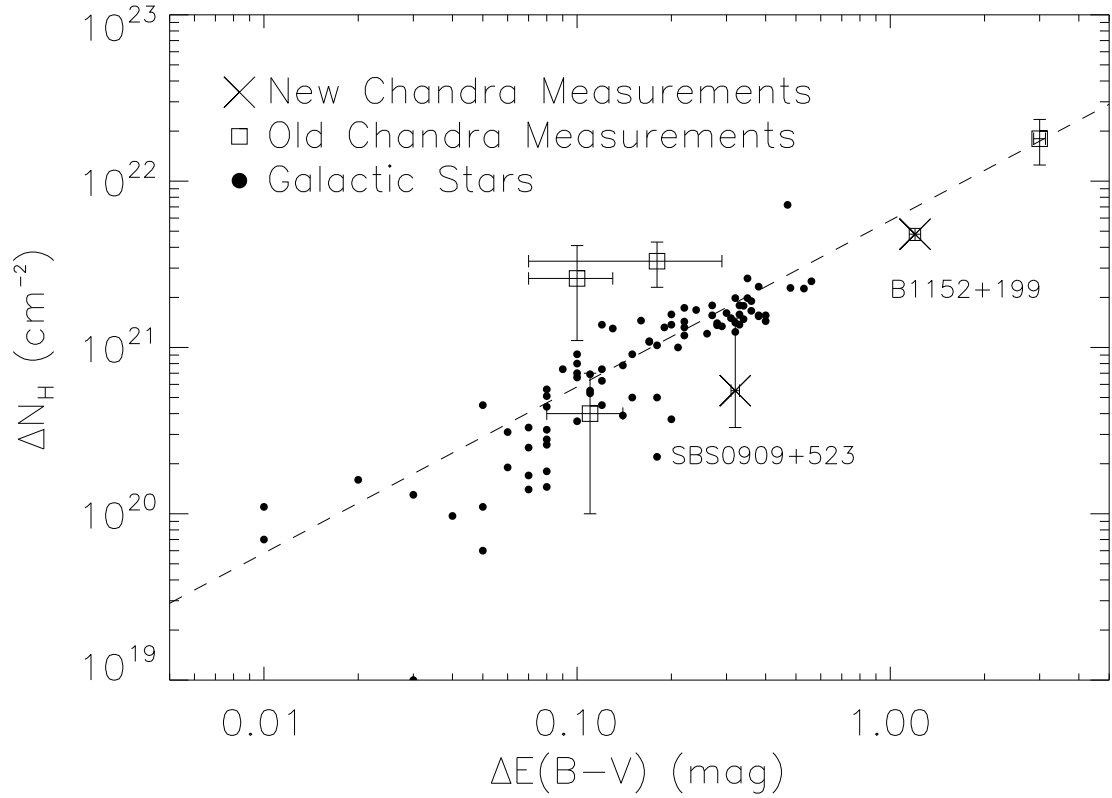}
\caption{$\Delta \nh$ versus $\Delta E(B-V)$ for the gravitational lenses (squares and crosses) and \nh\ versus $E(B-V)$ for Galactic stars from Bohlin \etal (1978, filled circles).  The dashed line shows the mean Galactic relation.  The two crosses are the new \chandra\ measurements for \sbs\ and \bone. \label{fig:dtg}}
\end{figure}

\begin{figure}
\epsscale{1}
\plotone{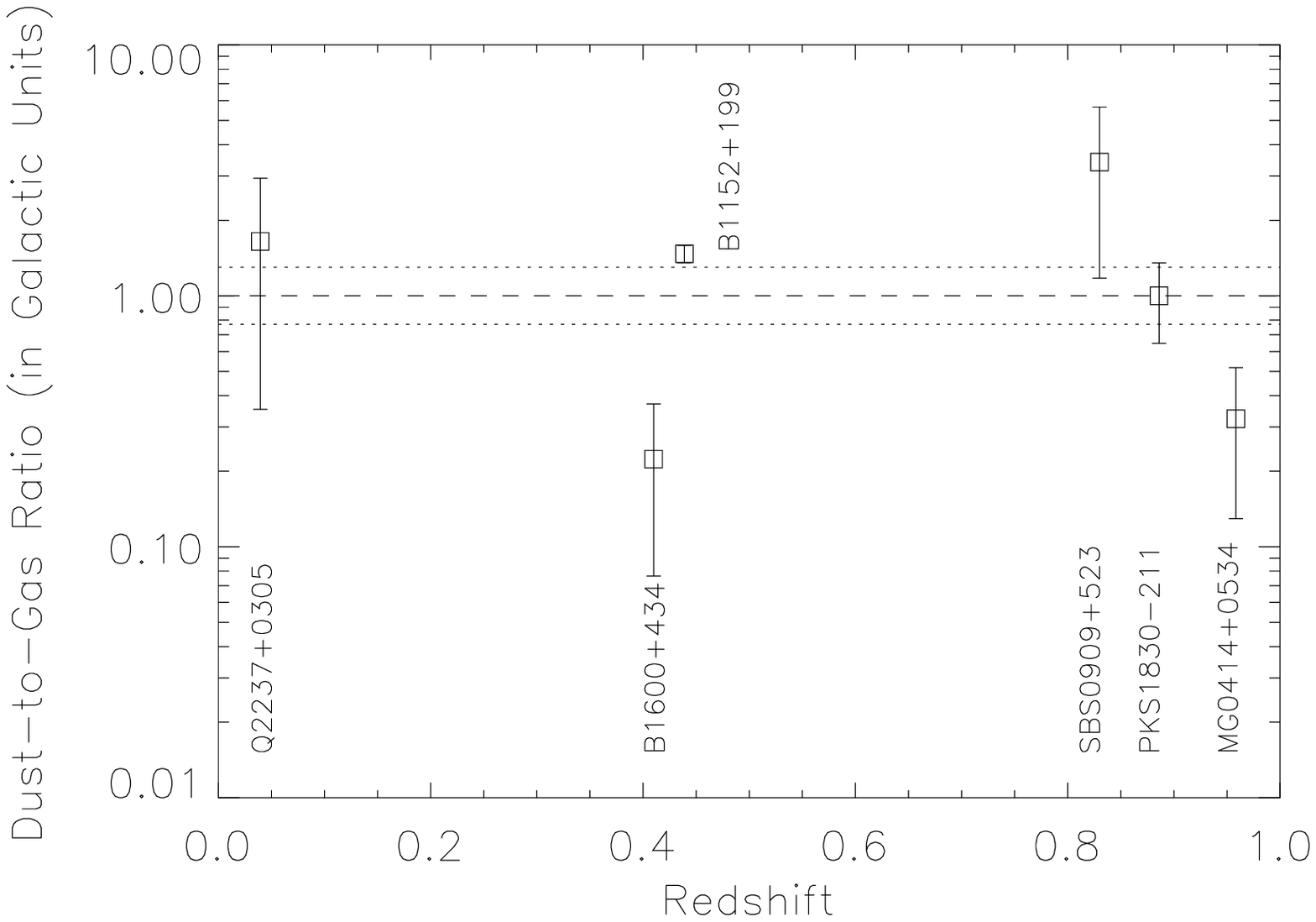}
\caption{The the dust-to-gas ratio (squares) versus redshift 
in Galactic units of 
$1.7\times 10^{-22}$ mag cm$^2$ atoms$^{-1}$.  The dashed and dotted lines show the Galactic value and its intrinsic scatter of 30\%.\label{fig:zevo}}
\end{figure}

\section{Discussion}
We summarize all available differential absorption measurements in Table~\ref{tab:dtg},
including the three results from this paper (\sbs, \fbqs, and \bone) and the 
five existing measurements from Dai et al.\ (2006).  
We find an average dust-to-gas ratio of
\begin{equation}
E(B-V)/{N}_{\rm H} = (1.5\pm0.5)\times 10^{-22}{\rm mag~cm}^2~{\rm atoms}^{-1}
\end{equation} 
with an intrinsic scatter of $\simeq 40\%$, where we must exclude \fbqs\ and \he\ from the fits.  
If we allow for no intrinsic scatter, we
found a ratio of $(2.3\pm0.2) \times 10^{-22}$ mag cm$^2$ atoms$^{-1}$ with
$\chi^2/N_{dof}= 3.0$, indicating the existence of the additional scatter
we include in our standard estimate.
These estimates are consistent with our previous estimate in Dai et al.\ (2006) and the average Galactic value of $1.7\times 10^{-22}$ mag cm$^2$ atoms$^{-1}$ \citep{bsd78}.
The estimates for the intrinsic scatter are also consistent with local estimates given our small sample size.
We compare the results for the lens galaxies to the local stellar sample 
from \citet{bsd78} in Figure~\ref{fig:dtg}, 
and plot the evolution of dust-to-gas ratios in Figure~\ref{fig:zevo}.
Contini \& Contini (2007) examined this question by modeling the bulk emission of luminous
IR galaxies, finding that the dust-to-gas ratio is significantly higher than the Galactic value in star forming and central nucleus regions (see also Maiolino et al. 2001a,b) and consistent with the Galactic value elsewhere.
However, the scatter of their results is significantly larger than our more direct measurements.
In addition, Menard \& Chelouche (2008) found several $MgII$ absorbers at
$z\sim1$, which also have dust-to-gas ratios close to the Galactic value.
In our present results, we see no evidence for evolution in the dust-to-gas ratio to $z \simeq 1$, and with a larger sample we will
be able to test the ISM evolution models
of simulations \citep[e.g.,][]{dw98,ed01,in03}.

The primary systematic uncertainty for our results comes from 
chromatic microlensing of the quasar by the stars in the lens galaxy 
distorting the
optical and X-ray spectra.  Given the available information for the
systems we consider, it is possible that \he\ and \fbqs\ are affected 
by this problem while the rest show no evidence of a problem.  In addition,
since we have confined
our studies to systems with significant evidence for extinction, 
we are relatively safe from minor chromatic microlensing.
If microlensing is creating the large color differences between the 
images then we should also see rapid changes in the colors because a strong
chromatic effect requires the source to lie near or on a caustic.
Nonetheless, it is possible from two of the systems (\he\ and \fbqs) 
that microlensing
can mimic extinctions of order $\Delta E(B-V)=0.1$ mag.  However, that the
lenses follow the local dust-to-gas ratio and with similar scatter
is strong evidence that the remaining systems are little affected 
by high levels of chromatic microlensing effect.
Monitoring these systems at both optical and X-ray 
wavelengths would
allow secure determination of the absorption properties while simultaneously
allowing us to determine the spatial structure
of the source quasar.

\acknowledgements
We thank {\'A.} El{\'{\i}}asd{\'o}ttir, B. Menard, L. J. Goicoechea, and the anonymous referee
for helpful comments.  We gratefully acknowledge the financial support by CXC grant G07-8080.

\end{document}